# Chemical and Morphological Transformations of a Ag-Cu Nanocatalyst During $CO_2$ Reduction Reaction

Gustavo Zottis Girotto,[1,2] Maximilian Jaugstetter,[3] Dongwoo Kim,[2,4] Ruan M. Martins,[5] André R. Muniz,[5] Miquel Salmeron,[3] Slavomir Nemsak,[2,6,*] Fabiano Bernardi[1,*]

[1] Programa de Pós-Graduação em Física, Instituto de Física, Universidade Federal do Rio Grande do Sul, Porto Alegre, RS, Brazil

[2] Advanced Light Source, Lawrence Berkeley National Laboratory, Berkeley, CA, USA

[3] Materials Science Division, Lawrence Berkeley National Laboratory, Berkeley, CA, USA

[4] Department of Physics and Photon Science, Gwangju Institute for Science and Technology, Gwangju, South Korea

[5] Department of Chemical Engineering, Universidade Federal do Rio Grande do Sul, Porto Alegre, RS, Brazil

[6] Department of Physics and Astronomy, University of California, Davis, CA, USA

**Abstract**

The conversion of $CO_2$ into high-value chemicals through a photoreduction reaction in water is a promising route to reduce the dependence on fossil fuels. Ag nanoparticles can drive this reaction via localized surface plasmon resonance, but their low selectivity limits usage in industry. Enhancing selectivity toward hydrocarbons or alcohols requires addition of a co-catalyst such as Cu. However, the stabilized surface state created by Ag-Cu interactions is still poorly understood. In this work, soft x-ray Ambient-Pressure X-ray Photoelectron Spectroscopy (AP-XPS) and Grazing-Incidence X-ray Scattering (AP-GIXS) were used to investigate the evolution of Ag-Cu nanoparticles under $CO_2$RR-like conditions. AP-XPS revealed Ag and Cu surface and sub-surface diffusion, while AP-GIXS tracked change of shape and size of nanoparticles induced by diffusion mechanics. Under 532 nm laser irradiation, further oxidation of Cu and Ag sub-surface diffusion were observed, providing invaluable insights into the dynamic restructuring of the catalyst under reaction conditions.

**Ag-Cu materials as promising catalysts for CO$_2$RR**

The CO$_2$ reduction reaction (CO$_2$RR) is an ambitious strategy to overcome the current energy needs and global warming issue [1]. One possibility is to combine CO$_2$ and H$_2$O to drive the CO$_2$ photoreduction reaction, which is also known as artificial photosynthesis [2]. Metals and metal oxides have been investigated in the reaction to produce alcohols and hydrocarbons [2-4]. In particular, it has been demonstrated that Au [5] and Ag [6] nanoparticles can initiate CO$_2$ reduction using H$_2$O as a proton source under only visible light exposure. As compared to purely electrochemical CO$_2$ reduction, light irradiation shifts the onset of CO formation on Ag electrodes to less cathodic potentials and increases the reaction rate [7]. However, while noble metals are able to activate the reaction, they exhibit insufficient selectivity [6]. Moreover, it has been demonstrated that Cu allows controlling selectivity, which is dependent on its oxidation state, despite its low activity for artificial photosynthesis reactions [8]. Several studies have previously taken advantage of the combination of Cu and Ag to control activity and selectivity in electrochemical CO$_2$RR [9-11] but the same is not found for CO$_2$ photoreduction. One of the few studies that used the Ag-Cu system exploiting visible light with dendritic photoelectrodes observed selectivity towards acetate at low overpotentials [12], not achievable with electrochemical CO$_2$RR only.

Cu presents a high adsorption energy to intermediate products such as *COOH, which can be transferred to regions of low concentration of adsorbed species such as the Ag surface [13]. In contrast, increasing the Ag concentration of these alloys lowers the average binding energy of adsorbed *CO species [14]. The selectivity for CO$_2$RR is improved by increasing the Ag-Cu interfacial area [13]. Interestingly, competitive hydrogen evolution reaction has been observed to be suppressed by the presence of Ag at the surface [15], but enhancement has also been claimed to be derived from a charge transfer effect [16]. Charge transfer was also suggested to strengthen the adsorption of *CO intermediates on Cu sites after CO$_2$ adsorption on Ag sites [17], and electron-deficient Cu sites are hypothesized to provide adsorption sites for alcohol product intermediates [18]. There are reports in the literature about atomic configuration evolution of Ag-Cu nanoparticles during CO$_2$RR. Single atom Ag incorporation into Cu can dynamically relocate during CO$_2$RR, promoting rich *CO regions on the surface [19]. Moreover, a stabilized Cu$^+$ overlayer was also found to be induced by Ag presence during electrochemical CO$_2$RR with electrodeposited AgCu alloys [20].

In parallel to the performance of the Ag-Cu system in CO$_2$RR, some studies analyzed chemical and morphological changes occurring in this system under different conditions.

Thin $Cu_2O$ shells around Ag have been observed to present poor stability during catalytic reactions due to a large strain at the Ag/Cu interface [21], where Cu dewetting produces small clusters separated from the Ag nanoparticles [16]. Furthermore, phase separation of Ag-Cu occurs during electrochemical $CO_2RR$ while Cu suffers reduction with cathodic currents [22]. Ag shells can be grown on Cu(0) nanoparticles by galvanic exchange method [23], but they may age towards dewetted Ag nanoparticles due to local stress [24], or be displaced to the core at the same time Cu shows gradual oxidation [25]. Density Functional Theory calculations have also shown that dilute Cu atoms on Ag slabs tend to remain at the sub-surface, but an oxidative environment at high temperature induces the formation of Cu-O at the surface [26], where a metastable configuration was recently observed [27].

The oxidation state and morphology of Cu nanoparticles evolve during photoreaction due to photocorrosion [28], where either oxidation or reduction processes occur. The mixing with Ag leads to another complexity in this regard, and there is a poor understanding about possible changes on the atomic arrangement and morphology of Ag-Cu nanostructures during photochemical $CO_2RR$. Determining the surface atomic population and nanoparticle morphology during reaction conditions is of utmost importance to further predict the correct reaction mechanisms. Consequently, it allows the design of future improved photocatalysts for $CO_2RR$. In this study, we apply a multi-modal approach through Ambient-Pressure X-ray Photoelectron Spectroscopy (AP-XPS), Grazing-Incidence X-ray Scattering (AP-GIXS), and in situ X-ray Absorption Spectroscopy (in situ XAS) measurements to elucidate the behavior of Ag-Cu nanoparticles during photochemical $CO_2RR$. *Ex situ* microscopy measurements provide valuable information on the as-prepared and post-mortem samples' morphology. Molecular dynamics simulations were also carried out to help in the interpretation of the experimental findings.

### *Ex situ* characterization

A typical SEM image of the Ag-Cu nanoparticles before and after exposure to $CO_2RR$ conditions is displayed in Figure 1(a) and 1(b). The as-prepared Ag-Cu nanoparticles present a mean diameter around 7 nm. SEM measurements confirm that nanoparticles' lateral dimensions slightly increase after $CO_2RR$ (Figure 1(c)). The analysis of AFM images (Figure 1(d) and Figure S1 of the SI) demonstrates that the particles show slight elongation in the direction normal to the surface. The UV-Vis spectra (Figure 1(e)) of the Ag-Cu nanoparticles show an increase in background as compared to the bare Si case. The increase in reflectance centered at 450 nm evidences the existence of Local Surface Plasmon Resonance (LSPR) due to Ag nanoparticles, which may promote the generation of hot carriers [29] or increase the electromagnetic near-field [30] that induces $CO_2RR$ [31]. After exposure to $CO_2RR$, the spectrum is clearly modified. The inset shows that a new oscillation around 550 nm arises, thus close to the wavelength of the laser irradiated to activate the reaction (532 nm), which is consistent with some structural modification taking place. This is elucidated further below by analysis of AP-GIXS data.

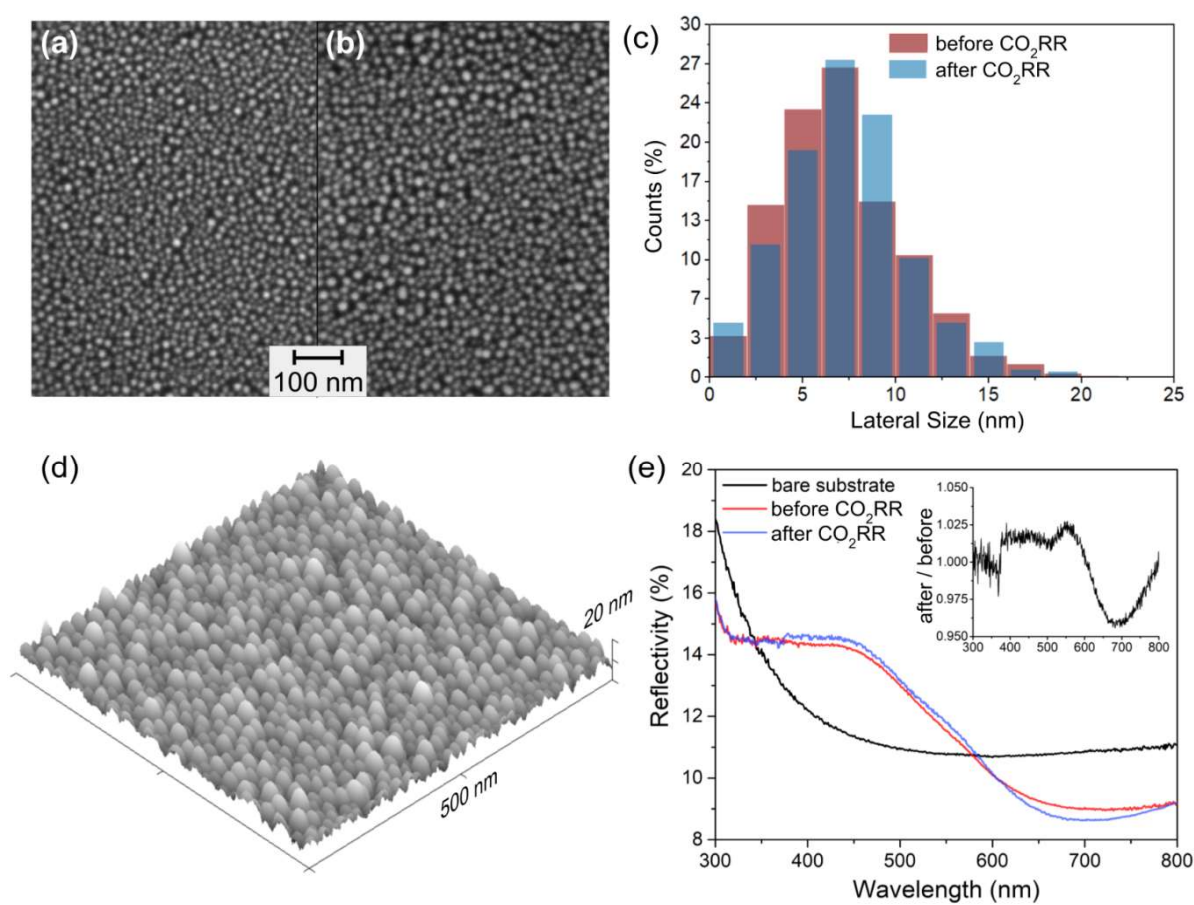

Figure 1: Typical SEM images of AgCu nanoparticles (a) before and (b) after exposure to $CO_2RR$. (c) Histogram of nanoparticle lateral width before and after exposure to $CO_2RR$. (d) Typical AFM image obtained of the sample after exposure to $CO_2RR$. (e) UV-Visible spectroscopy measured in total reflectance mode before and after exposure to $CO_2RR$. The inset shows the ratio between the spectra of the sample after and before exposure.

*In situ* **characterization of the nanoparticles surface state**

The AP-XPS measurements of energy regions relevant for the sample (Cu 3p, Ag 4p, Ag 3d, C 1s, and O 1s) using 695 eV excitation energy are displayed in Figure 2. The Cu 3p region shows contributions from $Cu^+/Cu(0)$ and $Cu^{2+}$ [32] with an energy separation of 1.5 eV. Similar components are observed using 1240 eV photon energy (see Figure S2 of the SI). The CuO [33] or $Cu(OH)_2$ [34] species possess an average separation of around 1 eV and 2 eV to the Cu(0) component, respectively. The Cu $2p_{3/2}$ XPS spectrum of the sample measured using an Al Kα source (Figure S3 of the SI) indicates a 2 eV separation between $Cu^{2+}$ and $Cu^+/Cu(0)$ components, which is consistent with the presence of $Cu(OH)_2$ at the surface [34]. The high concentration of $Cu^{2+}$ at the surface is expected since the sample has been exposed to air and humidity before being introduced to the experimental chamber. Only metallic Ag(0) is found at Ag 3d XPS region of the as prepared sample [35]. The C 1s region exhibits a large concentration of carbonate ($CO_3^{2-}$) species at binding energy around 289.5 eV [36], originating from the contamination during the growth in the evaporation chamber and the subsequent air exposure. The O 1s region presents two components associated to $SiO_x$ and $CuO_x$ (convolution of $Cu^{2+}$ and $Cu^+$ species) [37] related to the substrate and Ag-Cu nanoparticles, respectively. The normalized Cu/Ag ratio obtained using the Cu 3p and Ag 4p areas is 1.8, and the one obtained with 1240 eV photon energy is around 1.6 (Figure S4 of the SI). Simulation with the Electron Spectra for Surface Analysis (SESSA) [38] software was used to extract the relative concentrations of Cu and Ag from the data collected at 695 eV. Figure S5 of the SI shows that a spheroid with a 6-7 nm Ag core diameter covered by a 1 nm CuO shell reproduces adequately the relative intensity measured in the XPS spectra. Thus, the representative Ag-Cu stoichiometry is approximated as Ag1Cu1.

The annealing process at 200 °C removes the $CO_3^{2-}$ component in the C 1s region, and the $Cu^{2+}$ component almost disappears, indicating that Cu has reduced to either $Cu^+$ or Cu(0). This observation correlates with the decrease of $CuO_x$-related components in the O 1s region. A dramatic decrease of the normalized Cu 3p/Ag 4p intensity ratio from 1.8 to 0.4 during $H_2$ exposure implies that the concentration of Ag atoms at the surface increases. The subsequent introduction of a 40 mTorr $CO_2$ + 40 mTorr $H_2O$ atmosphere at RT results in a new low-binding energy component in the Ag 3d region, interpreted as $AgO_x$ [35]. Furthermore, another component in the O 1s region emerges at about 532 eV. This component is linked to the oxidation of Ag, and it is simultaneously observed as a relative increase of the $Cu^{2+}$ component in Cu 3p region. The dissociation of $CO_2$ at the surface generates surface O and C-O species (evidenced by C-O component increase in C 1s spectra). These results point to the existence of interfacial Cu-Ag-O chemical species [39].

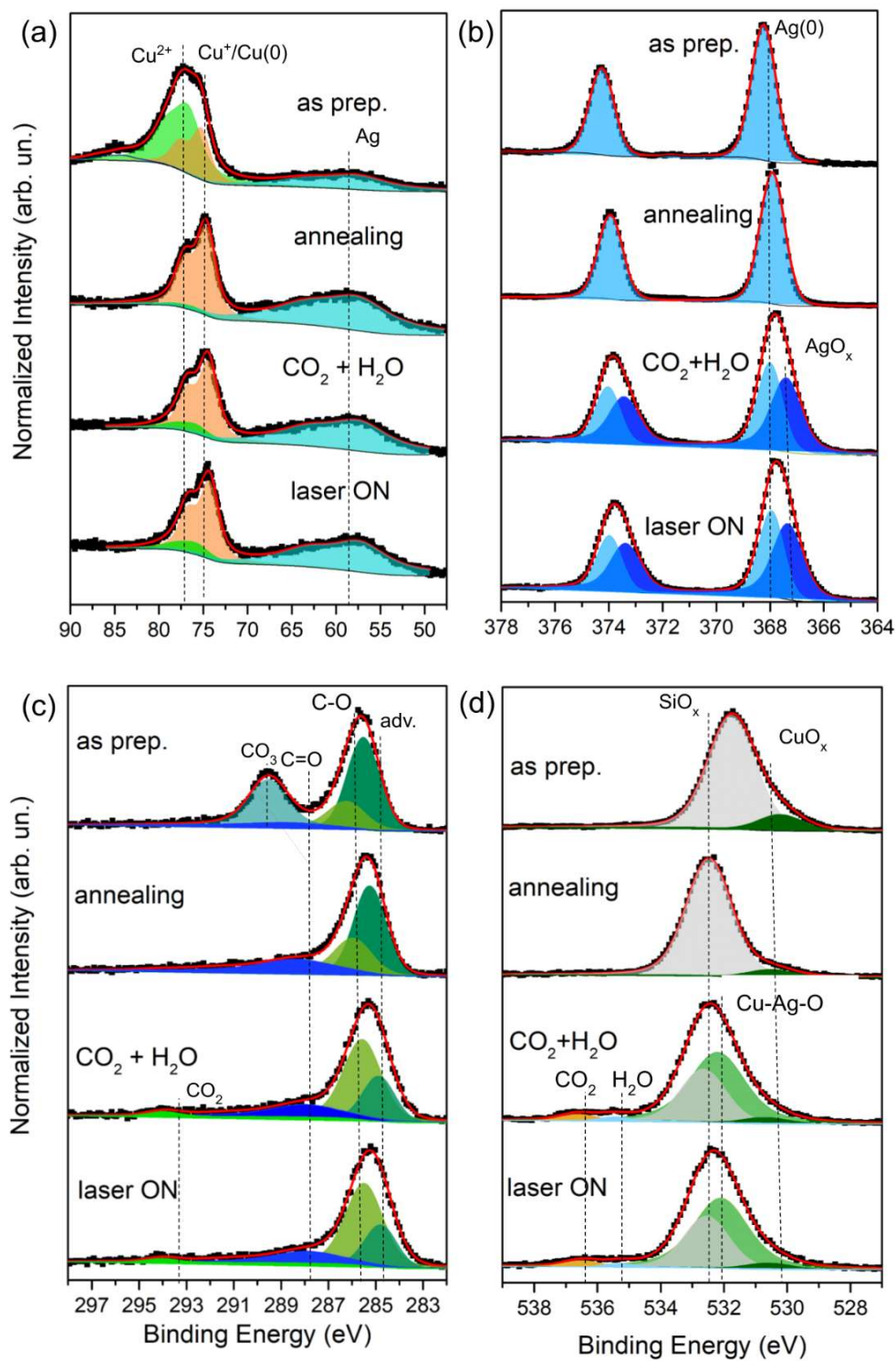

Figure 2: AP-XPS spectra at (a) Cu 3p + Ag 4p, (b) Ag 3d, (c) C 1s, and (d) O 1s electronic regions in the as-prepared, after annealing at 200 °C in 20 mTorr $H_2$, during $CO_2$RR with 40 mTorr $CO_2$ + 40 mTorr $H_2O$ with laser off, and during 532 nm laser on conditions.

Figure 3(a) displays how four distinct parameters change after each exposure condition. The Cu/Ag ratio increases after dosing $CO_2$ and $H_2O$. Although the $AgO_x$ percentage increases with $CO_2$ and $H_2O$ dosing, it does not change after laser irradiation. The Cu-Ag-O component normalized to total C content increases following $CO_2$ + $H_2O$ dosing, but then it decreases after laser irradiation. The C-O concentration normalized to total C content increases after every condition. Therefore, the laser triggers substitution of surface Ag by the Cu atoms, as indicated by the decrease of the Cu-Ag-O signal, which suggests Ag-Cu bond breaking. However, the constant $AgO_x$/Ag evidences that Ag is not further oxidized by this last condition. Similar behavior was found during the AP-XPS measurements performed at 1240 eV excitation energy (Figure S6 of the SI), and for a Ag-Cu powder system synthesized through a precipitation method (short discussion in the end of the SI). Furthermore, annealing the sample in an $O_2$ atmosphere promotes the complete substitution of the Ag atoms at the surface by a $Cu^+$ layer (Figure S7 of the SI), which is predicted by the lower surface energy of $CuO_x$ relative to Ag [25].

It has been suggested [39] that $CO_2$ adsorption at the surface of Ag-Cu alloys requires the presence of surface O, which is observed in the low binding energy component at O 1s. It cannot be excluded that the C-O and C=O components in C 1s spectra, which are still present after annealing, may participate in this mechanism. The growth of C-O component is followed by the decrease of C=O and adventitious C after dosing $CO_2$ and $H_2O$ (Figure S6 of the SI). The interaction of the surface carbon with $CO_2$ + $H_2O$ induces the exchange or conversion of the C-C or C-H species to C-O and/or the volatilization. However, with the laser turned on, the C-O increase is not followed by the further decrease of adventitious C, indicating a different mechanism. The laser irradiation must be another trigger to further promote the dissociation of $CO_2$ at the surface, which is associated with the substitution of Ag for Cu at the surface.

*In situ* XAS spectra measured in total electron yield mode during the different conditions at Cu $L_3$ edge are presented in Figure 3(b). The as-prepared condition shows two different features: the one around 930 eV is expected due to the $Cu^{2+}$ species at the surface, while the one near 934 eV corresponds to the $Cu^+$ species [40,41]. These two features agree qualitatively with AP-XPS measurements. After annealing, the spectrum changes dramatically: the $Cu^{2+}$ feature disappears, and the oscillation at around 937 eV evidences the appearance of a Cu(0) component [41], which could not be resolved with AP-XPS only due to binding energy overlap of $Cu^+$ and Cu(0) components. It has been shown that the segregation behavior of Ag to the surface in metallic Ag-Cu alloy nanoparticles is energetically preferred [42], as will also be supported by our MD simulations described below. After dosing $CO_2$ + $H_2O$, the shoulder at 937 eV decreases in intensity, and the white line narrows, evidencing Cu(0) oxidation to $Cu^+$. Cu oxidation starts before laser irradiation, but the incidence of visible light further accelerates the process. The stabilization of $Cu^+$ at the surface was also observed during electrochemical $CO_2RR$ due to interaction with *CO [20]. A detailed comparison of XAS measured in laser off and laser on conditions (Figure S8 of the SI) further shows fine differences at the pre-edge and post-edge regions, representing changes in oxidation state.

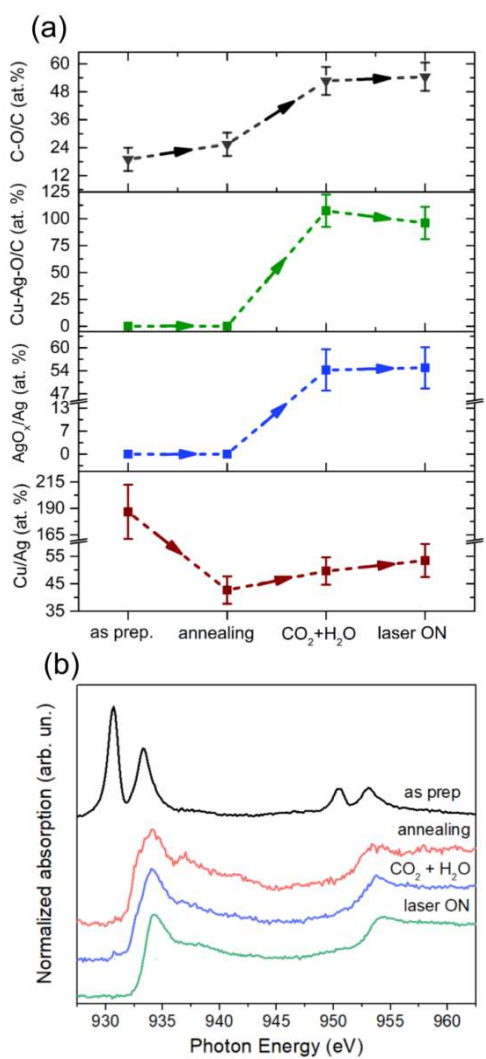

Figure 3: (a) correlations taken out of AP-XPS measurements between Cu/Ag, $AgO_x$/Ag, Cu-Ag-O/C, and C-O/C ratio on the surface obtained during different conditions. (b) XAS measurements measured in drain-current mode in the same conditions.

**In situ characterization of the nanoparticles morphology**

AP-GIXS scattering patterns measured under the same $CO_2$RR conditions are shown in Figure 4. The as-prepared sample (Figure 4(a)) exhibits broad, intense features around $q_y$ = 0.35 nm$^{-1}$, indicating a strong structure factor affecting the in-plane scattering components. After annealing (Figure 4(b)), two major changes occur: 1. the structure factor maxima shifts towards the origin, suggesting increased separation between the particles' center of mass; 2. an additional out-of-plane feature appears. The extension of the oscillation above the beamstop footprint indicates a decrease in particle height. Similar trends emerge after $CO_2$ and $H_2O$ exposure (Figure 4(c)), with further shift of the maxima (Figure S9 of the SI) and increased out-of-plane oscillation length, suggesting that particles are growing laterally and their elongation along sample normal is decreasing. Further evidence comes from the Si 2p signal decrease in survey spectra (Figure S10 of the SI), supporting the interpretation of scattering data of particles spreading and covering the substrate. We recently demonstrated that monometallic Cu spreads over the support even at relatively low temperatures [43].

Linecuts at around $q_y$ = -0.3 nm$^{-1}$ were taken to perform a 1D fitting procedure considering the Boucher sphere model present in the SASfit package [44], which is based on a study of alloy particles [45]. Even though the real nanoparticles are not exactly spherical, the simplicity of the model properly captures the main expected trends. The fitting takes into account polydispersity by considering a log-normal size distribution and a variable scattering contrast along the radius of the nanoparticles based on a single parameter (fitted data are shown in Figures S11 and S12 of the SI). The scattering contrast as a function of the radial distance from the center to the surface of the nanoparticle (with radius equal to the average size of the particle distribution) obtained after the fitting are shown in Figure 4(d). First, the as-prepared condition shows a scattering contrast that decreases as the radial distance increases, supporting the Ag-rich core and Cu-rich shell hypothesis. After the annealing procedure, the scattering contrast at the surface is higher than at the core, and the external radius (particle height) shrinks, evidencing that Ag is replacing the Cu atoms at the surface, in a full agreement with the AP-XPS results. After $CO_2$ + $H_2O$ exposure, the scattering contrast has a smaller intensity, which is related to the overall oxidation of the nanoparticles.

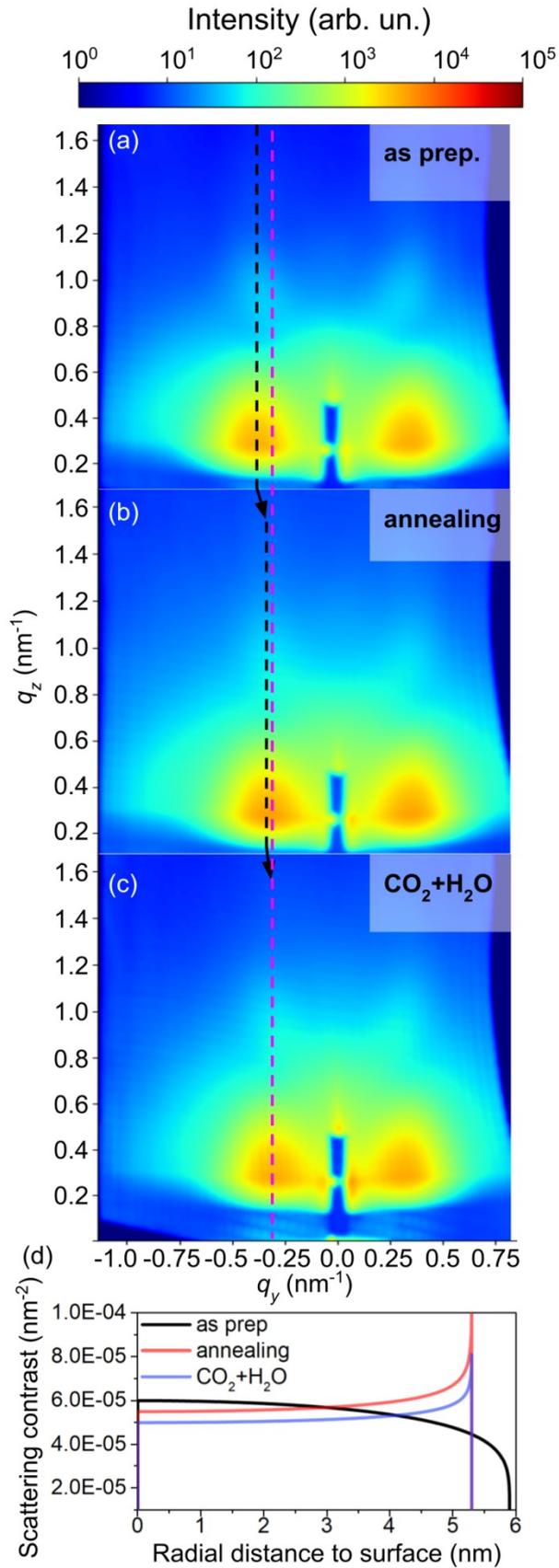

Figure 4: (a)-(c) AP-GIXS measurements of the sample as prepared, after annealing, and upon 40 mTorr $CO_2$ + 40 mTorr $H_2O$ exposure. Black lines are placed as a guide to the eye

centered at the high intensity lobe, while the magenta line shows where linecuts were taken for further analysis. (d) Scattering contrast as a function of the radial distance to the surface of the average size nanoparticle, obtained after fitting a linecut with the Boucher sphere model.

A simulation of the scattering pattern with BornAgain software package [46] (Figure 5(a)) was performed in order to verify the overall morphology of the particles present in the as prepared sample. The simulation takes into account a spheroid form factor composed of core-shell particle geometry, depicted in Figure 5(b). The core and shell regions are approximated by the optical constants of Ag(0) and CuO, respectively [47]. A recent study at beamline 11.0.2 showed that the GIXS setup enables obtaining a good estimative for shell thickness [48]. The estimated core diameter is of 6 nm, while the shell thickness is of 1 nm, similar to what was previously estimated using AP-XPS (SESSA simulation) and AFM measurements. The height of the spheroid is ca. 10 nm, with 20% size polydispersity. The as-prepared sample's interparticle spacing histogram from SEM images shows a maximum in the range of 3 – 4 nm (Figure S13 of the SI). This is considered in the fitting by mixing a paracrystal function with an average distance between the centers of particles of 11 nm. A linecut taken along $q_y$ = -0.3 nm$^{-1}$ in the simulated BornAgain pattern fitted to the Boucher model (Figure S14 of the SI) shows a good agreement with the fitting of the as-prepared sample (Figure 4a), thus further validating our model.

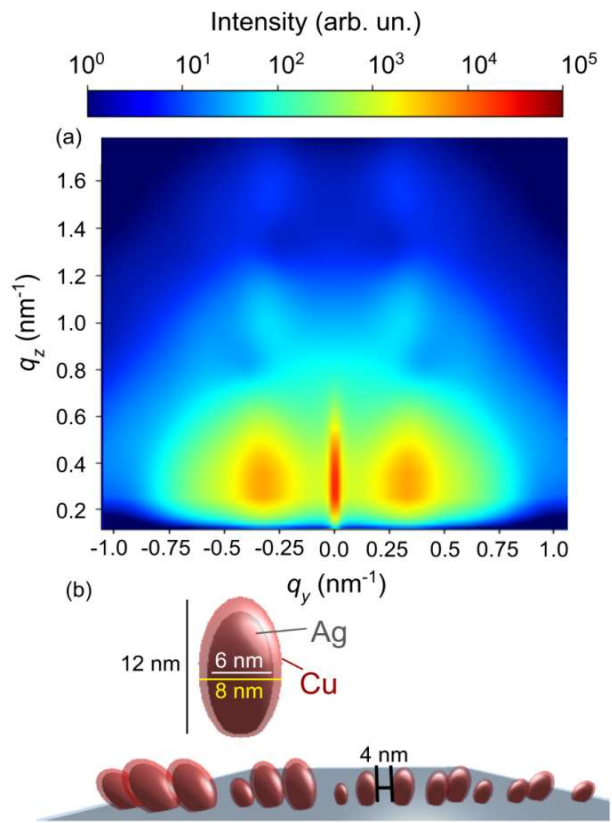

Figure 5: (a) Simulated scattering pattern of the as-prepared sample using BornAgain software, and (b) the nanoparticle configuration used for the simulation.

After laser irradiation, resonant AP-GIXS captured at the Cu $L_3$ edge was measured in order to estimate the Cu distribution within the nanoparticles (Figure 6(a)). The pattern at off-resonance (below the absorption edge) is still heavily influenced by the structure factor contribution. However, at the Cu $L_3$ edge white line (934.2 eV), the structure factor weakens as the Cu scattering contribution is suppressed. The smearing of the structure factor is due to a weakening of the hard-sphere condition, which is embedded into the paracrystal lattice. This happens because the surface is primarily composed of Cu, imposing the limiting condition of particle-particle separation. It is important to notice that Ag-Cu scattering cross terms are present [49], which limits a complete reconstruction with only two different incident energies. After laser irradiation at the off-resonance condition, the average external radius is shrunk (Figure 6(b)) as compared to the case before laser irradiation, but the shape is still similar to the previous cases. The fitting of the resonant condition enables identifying that the increase of the scattering contrast at a smaller external shell represents Ag located not at the surface anymore but at the sub-surface region. Again, it supports the previous observation by AP-XPS that the laser triggers the replacement of Ag by Cu directly at the surface. The core region still shows a smaller scattering contrast, indicating that the core is composed of an Ag and Cu mixed phase.

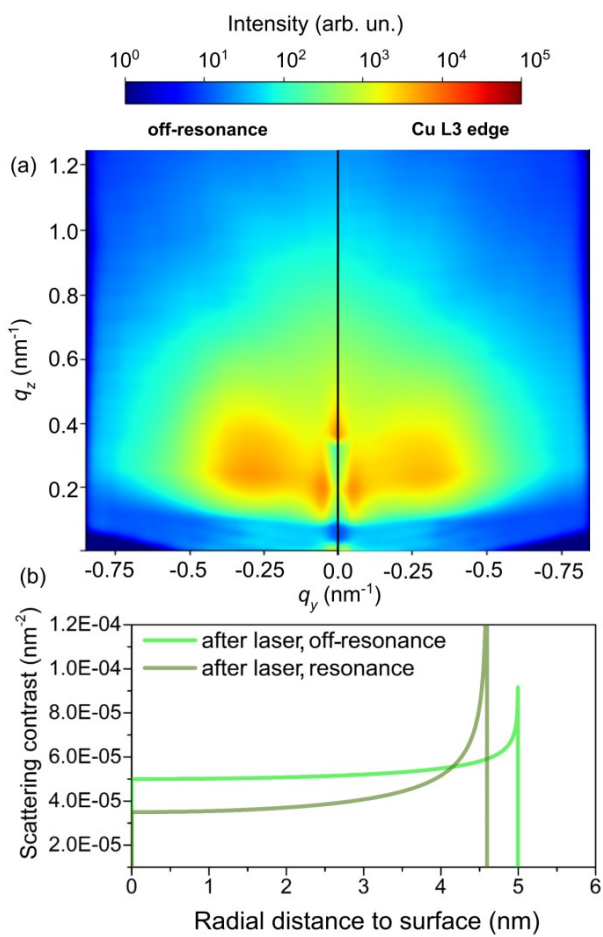

Figure 6: (a) Comparison between scattering patterns obtained after 532 nm laser irradiation around Cu $L_3$ edge, in off-resonance and resonance conditions. (b) Scattering contrast as a function of the radial distance to the surface of the average size nanoparticle obtained after fitting a linecut with the Boucher sphere model.

MD simulations of a thermal annealing process were performed on an initially elongated core-shell nanoparticle with an approximate Ag(0):Cu(0) 1:1 stoichiometry, which are displayed in Figure 7(a) (equivalent images for nanoparticles with varied compositions are shown in Figure S15 of the SI). This system is equivalent to the experimental one after annealing in $H_2$ atmosphere used to reduce $Cu^{2+}/Cu^+$ to Cu(0). A general pattern was seen for all nanoparticles regardless of composition, with the Ag atoms replacing the Cu atoms at the surface and the particles becoming more spherical as the temperature is increased (1000-1500

K). Annealing at 1000 K decreases the specific energy of the system (see Figure S16 of the SI), promoting the most stable atomic configurations during the simulation. Because of a significant tendency for Ag segregation, Cu atoms are found in internal pockets even at the highest temperature of 1500 K. This results in the formation of multiple interfaces between Ag and Cu inside the nanoparticle, which increases the specific energy of the system (Figure S16 of the SI). Indeed, we recently discovered the formation of internal pockets in bimetallic nanoparticles for a similar system of Ni-Pd nanoparticles where Pd pockets are formed under $H_2$ exposure [50]. As mentioned in the Methods section, these temperatures must be higher than those used in the experiments (473 K) to enable observing relevant structural transformations within the MD timescale (ps-ns instead of minutes), therefore their absolute values should not be taken for direct comparison purposes, and only the trends must be considered (i.e., analyze which transformations occur or do not occur as we increase the annealing temperature).

Figure 7(b) compares the surface Ag-Cu stoichiometry to the full stoichiometry of a particle, while Figure 7(c) compares the latter to the mean/maximum radius ratio of a particle. The data suggest that when a particle containing metallic Cu and Ag atoms is subjected to a thermal treatment, there exists a transition temperature (1000 K) at which the mobility of atoms becomes dominant. Diffusion of Cu atoms spontaneously induces changes in morphology. This is further supported by the observation that Ag particles without Cu require annealing at very high temperatures (1500 K) to achieve a significant change in their elongation (see Figure S17 of the SI). At lower annealing temperatures (300 K - 800 K), all models remain elongated along the main axis and feature a very low concentration of Ag on the surface. Reasonably, higher Cu content and thicker shells require higher energy input to overcome kinetic barriers. A plateau region is formed by the $Ag_{0.59}Cu_{0.41}$ and $Ag_{0.66}Cu_{0.34}$ models in both Ag surface concentration and elongation after annealing at 1000 K, suggesting that certain Cu/Ag ratios favor these morphological changes. After annealing at 1500 K, all particles become nearly spherical, with Ag surface concentration increasing to approximately 90%.

An overall representation of the surface composition, observed through AP-XPS, and morphological changes, through AP-GIXS, and corroborated by MD simulations is summarized in Figure 7(d). First, the as-prepared sample is composed of a Cu-rich shell and an Ag-rich core. After annealing, the Ag atoms diffuse to the surface region, and the particle begins to become more spherical and extend laterally over the substrate. During $CO_2 + H_2O$ exposure, Ag atoms at the surface become oxidized, and Cu atoms diffuse to the surface region to form a Cu-Ag-O interface. After the sample is irradiated with the 532 nm laser while being exposed to $CO_2 + H_2O$, the Cu atoms replace Ag atoms at the surface, and the Ag atoms are trapped at the sub-surface region. The active phase is formed by the Cu overlayer at the Ag-Cu interface, which is similar to observed in other catalysts such as Pd-Co [51]. Chang et al. have observed that a reoxidation/reduction of $Cu^+$/Cu in Ag-Cu nanowires by cycling anodic to cathodic potentials induces the atomic diffusion of the species and enhances $CH_4$ formation [52]. The observation that these light-induced phenomena promote a similar oxidation step as the anodic cycle should be leveraged in the design of a photoelectrocatalytic system with improved selectivity and efficiency. Moreover, the observation by Liu et al. that subsurface O depletion in Cu electrodes during typical electrochemical $CO_2$RR shifts the selectivity to $H_2$ [53] suggests that the stabilized surface $Cu^+$ during laser irradiation is preferred for enhancing the production of hydrocarbons or alcohols.

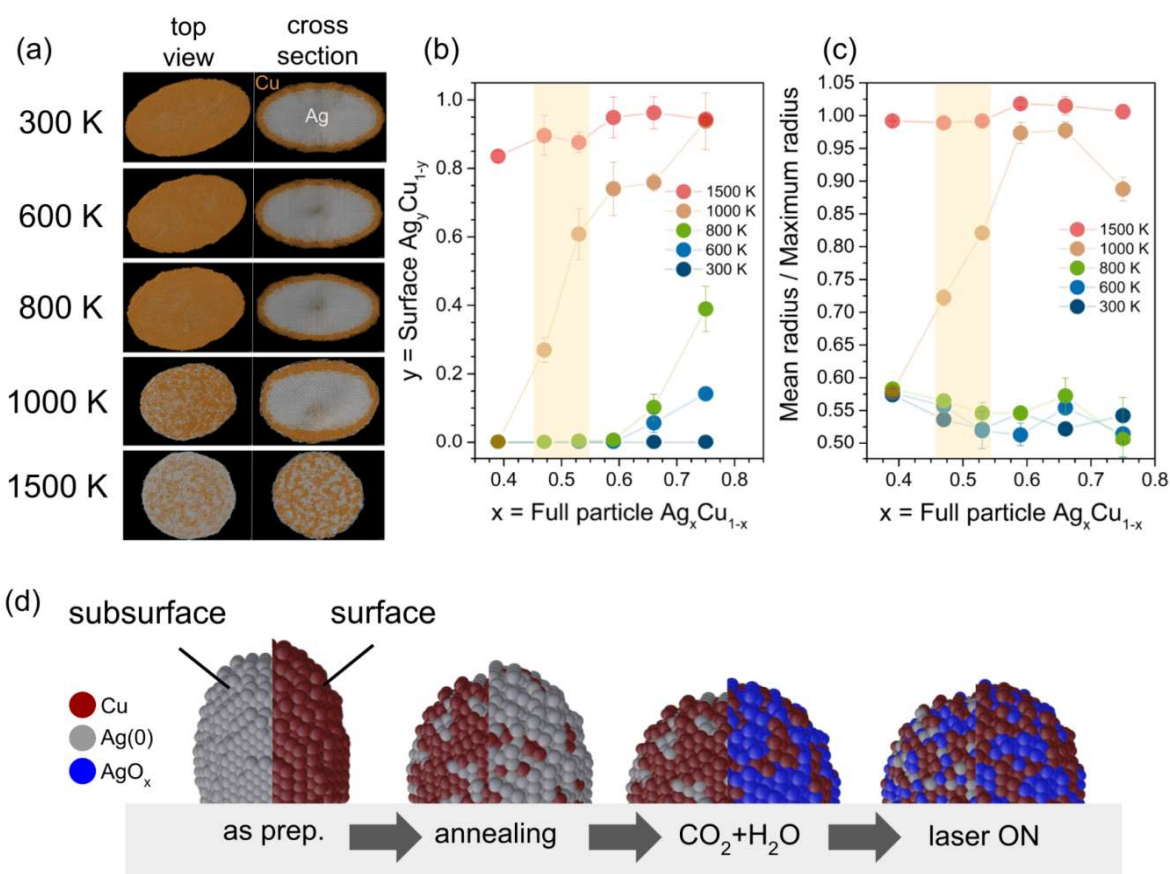

Figure 7: (a) Atomic configurations of core-shell 47% Ag - 53% Cu nanoparticles annealed at different temperatures in MD simulations. (b) Surface stoichiometry and (c) ratio of mean to maximum particle radius after thermal annealing as a function of the stoichiometry of the full nanoparticle. Shaded areas indicate points closest to experimental observation. (d) Schematic representation of the overall transformations observed during $CO_2$RR.

**Conclusions**

Ag-Cu nanoparticles were exposed to different conditions similar to those typical of the photochemical $CO_2RR$. The combination of AP-XPS and AP-GIXS measurements allowed capturing how Cu and Ag atoms reorganize inside the nanoparticles while they spread over the substrate when exposed to reaction-like conditions, which was corroborated by MD simulations. We observe the formation of a Cu-Ag-O interfacial phase during $CO_2$ + $H_2O$ exposure, accompanied by changes of nanoparticle's overall morphology during these processes. We further observed that visible light irradiation (532 nm) accelerates Cu(0) oxidation and Ag replacement at the surface during $CO_2RR$-like conditions, where Ag is trapped at the sub-surface. Our results suggest that this mechanism is responsible for enhanced activity and selectivity of the Ag-Cu system (see SI for discussion on reaction products). Correlating chemical and morphological transformations of this metal photocatalyst under *in situ* conditions was crucial to obtaining this fundamental understanding.

It is still an open question to which extent this is a purely photochemical induced transformation leading to charge transfer or whether other factors (such as the heat generated during the irradiation process) contribute. This can be disentangled by blue-shifting the irradiated light, which is planned for future studies. It will also be interesting to explore whether similar bimetallic systems exhibit comparable behavior, and whether selecting other substrates might change the observed effects. Furthermore, any kinetic effect related to the diffusion of Cu and Ag atoms is evidently hidden during this study because the measurements are not time-resolved. As the experimental and MD simulation data have shown, atomic diffusion is closely linked to morphology kinetics, and thus measuring time-resolved AP-GIXS and/or AP-XPS would provide data related to rate of diffusion. Finally, the findings can aid in the design of catalysts suited for cathodic electrochemical currents. Since Cu(0) is the stable phase during typical electrochemical $CO_2RR$, triggering oxidation via light exposure could enhance catalyst stability. Our results suggest that such catalyst materials could be dynamically controlled under light-off/light-on conditions to harness the diffusion mechanism and therefore stabilize the desired intermediates.

## Methods

The Ag-Cu nanoparticles were prepared on Si(111) p-type doped wafers using a commercial thermal evaporator system Denton DV-502A at the Molecular Foundry, Lawrence Berkeley National Laboratory (LBNL). The wafers were sonicated with isopropanol and water, followed by drying in $N_2$ gas flow. The evaporator system was evacuated using a scroll pump and a turbomolecular pump, reaching a stable pressure typically around $10^{-5}$ Torr after 2 h. Ag was first evaporated until a thickness of 2 nm was reached with a rate of 0.5 Å/s, measured using a quartz crystal microbalance. The sample plate was then heated for 2 h at 60 °C to outgas and then to 250 °C during 12 h to form Ag nanoparticles. Afterwards, the same procedure was used to evaporate 0.5 nm thickness of Cu which was annealed at 250 °C for another 12 h.

AP-XPS measurements were taken at beamline 9.3.2 at ALS [54], which is equipped with a VG-Scienta R4000 HiPP electron analyzer. All the measurements were performed using a 695 eV photon beam energy. Analyzer pass energy of 100 eV was used with step sizes of 0.1 eV and 0.5 eV for the high-resolution and survey spectra, respectively. The AP-XPS spectra were collected in the survey, Cu 3p, Ag 4p, Ag 3d, Si 2p, C1s, O1s, and valence band electronic regions. The sample was initially measured in the as-prepared condition, which was followed by dosing 20 mTorr $H_2$ and annealing to 200 °C. At this temperature, the sample remained 30 min and a second set of AP-XPS measurements were taken. After cooling down to RT in UHV, another set of measurements were taken. Then, 40 mTorr $CO_2$ + 40 mTorr $H_2O$ were dosed into the main chamber during 30 min. After measuring the sample in this condition, a 532 nm DDPS laser was coupled to a power supply operating at 3.0 V and 0.3 A. The laser irradiated through the nozzle of the analyzer over the sample. A temperature rise of 5 °C on the sample was measured after 5 min of laser irradiation with the aid of a thermocouple, stabilizing in this temperature during laser operation. This condition was kept during 30 min, and new AP-XPS measurements were conducted after this period with the sample exposed to $CO_2$ + $H_2O$ and the laser on.

AP-XPS, AP-GIXS, and *in situ* XAS measurements were performed at APPEXS endstation at 11.0.2 beamline at ALS [55,56]. The APPEXS endstation is equipped with a SPECS PHOIBOS 150 NAP electron analyzer and an Andor iKon-L CCD mounted on the biaxial quasi-spherical manipulator to collect the scattered X-rays. A 532 nm laser probe (Spectra Solutions, Inc.) was placed inside the main chamber and operated at 800 mW. The AP-XPS and AP-GIXS measurements were taken using a 1240 eV photon energy, corresponding to an X-ray wavelength of 1 nm. The measurements were performed in grazing incidence using a 1º angle relative to the surface, below the critical angle of Si substrate (approximately 1.2º). AP-XPS measurements were collected in the survey, Cu 3p, Ag 4p, Ag 3d, Si 2p, C1s, O1s, and valence band electronic regions. A step of 1 eV and 0.1 eV, and pass energy of 10 eV were used for the survey and high-resolution spectra, respectively. For AP-GIXS, the scattered X-rays are collected at ± 12° along in-plane and 24° out-of-plane directions. In situ XAS measurements were performed at Cu L edge, from 920 eV to 960 eV, in total electron yield/drain-current mode using thermocouple wires connected to the Si wafer electrically isolated from the sample holder. The signal was obtained using a pico-Amperimeter configured with a 20 nA/V sensitivity, with incremental steps of 0.15 eV and a collection time of 0.5 s per point. Two spectra were averaged for each different condition. The sequence of experimental conditions applied to the samples in APPEXS setup was the same as during AP-XPS measurements performed at beamline 9.3.2. A similar temperature rise was observed when the laser was turned on.

SEM images were obtained at Molecular Foundry-LBNL using a Zeiss Gemini Ultra-55 microscope. The samples were measured before and after the AP-XPS and AP-GIXS measurements. The SEM images were obtained by detecting secondary electrons. The images were analysed with ImageJ software by selecting an appropriate threshold and automatic detection of grains [57]. AFM measurements were conducted at the Imaging Facility of the Molecular Foundry-LBL using a Asylum Jupyter AFM, and a PEAKFORCE-HIRS-F-A tip from BrukerNano. The AFM was operated in tapping mode, with a drive amplitude of 2 mW, 100 mV setpoint voltage, scan rate of 0.75 Hz, and 256 x 256 resolution. The optical properties of the samples were studied using ultraviolet–visible (UV–Vis) total reflectance spectra at CEOMAT-UFRGS on a UV-VIS-NIR spectrophotometer Cary 5000 (Agilent) in the wavelength range of 300–800 nm using an integrating sphere (DRA - 1800) with a sphere diameter of 150 mm. Measurements were taken of the sample grown on top of the Si substrate before and after exposure to every condition at beamline 11.0.2, and of the bare, clean Si substrate.

KOLXPD software was used for the AP-XPS data analysis. For the fitting of the high-resolution spectra, a Shirley-type background was subtracted. Voigt functions were used, where the Gaussian width of the peaks was fixed at the same value for each electronic region and the same excitation energy. The Lorentzian width was fixed for the same chemical component in a given electronic region for all the measured conditions. The relative binding energy of each chemical component was also fixed during the procedure. AP-GIXS raw data treatment is described on the SI. BornAgain software was used to simulate the AP-GIXS scattering patterns. A spherical detector was considered, with a 1º grazing-incidence X-ray beam on the sample. Different in-plane and out-plane linecuts were taken in the measured data to compare with the simulated pattern. SASfit software was used to perform the fitting procedure of selected measured data linecuts. The fitting used a fixed constant background and a Boucher sphere form factor.

Classical molecular dynamics (MD) simulations were carried out using the LAMMPS package [58] to study the atomic structure of AgCu nanoparticles of varying diameters and compositions annealed at different temperatures. The interatomic interactions were described by the EAM (Embedded-Atom Method) potential [59], with a suitable parameterization for Ag and Cu [60], which describes well the structure of individual phases and its alloys. To further elucidate the effect of annealing temperature and atomic composition on the features of Ag-Cu nanoparticle surfaces, we created core-shell (Ag-Cu) ellipsoidal nanoparticles with characteristic dimensions of $12 \pm 1$ x $7 \pm 1$ x $7 \pm 1$ nm$^3$ (dimensions slightly vary, depending on the composition), with compositions of 25, 34, 41, 47, 53, and 61% Cu (keeping the total characteristic dimensions and varying the shell thickness, the number of atoms varies from 17400 to 33700). The initial systems were submitted to thermal annealing at different temperatures (300, 600, 800, 1000 and 1500 K) starting from 0.1 K, using heating and cooling stages of 500 ps and equilibration steps of 3 ns (with timesteps of 5-8 fs). Higher temperatures than those used in experiments are employed to accelerate structural transformations and observe relevant phenomena in the timescale accessed by MD (nanoseconds, instead of minutes in the real experiments). The temperature was controlled using the Nosé-Hoover thermostat. The system energy was monitored to ensure that the annealing time was long enough to promote a steady value. The final morphology, characteristic dimensions, and surface composition of the nanoparticles were analyzed after annealing. Further information is given in the SI.

**References**

1. Bushuyev, O. S., De Luna, P., Dinh, C. T., Tao, L., Saur, G., van de Lagemaat, J., Kelley, S. O. & Sargent, E. H., What should we make with CO2 and how can we make it? Joule 2, 825–832 (2018).
2. Verma, R., Belgamwar, R. & Polshettiwar, V., Plasmonic photocatalysis for CO2 conversion to chemicals and fuels. ACS Materials Letters 3, 574–598 (2021).

3. Liu, K., Qin, R. & Zheng, N., Insights into the interfacial effects in heterogeneous metal nanocatalysts toward selective hydrogenation. Journal of the American Chemical Society 143, 4483–4499 (2021).
4. Xiao, C. & Zhang, J., Architectural design for enhanced C2 product selectivity in electrochemical CO2 reduction using Cu-based catalysts: a review. ACS nano 15, 7975–8000 (2021).
5. Shangguan, W., Liu, Q., Wang, Y., Sun, N., Liu, Y., Zhao, R., Li, Y., Wang, C. & Zhao, J., Molecular-level insight into photocatalytic CO2 reduction with H2O over Au nanoparticles by interband transitions. Nature Communications 13, 3894 (2022).
6. Devasia, D., Wilson, A. J., Heo, J., Mohan, V. & Jain, P. K., A rich catalog of C–C bonded species formed in CO2 reduction on a plasmonic photocatalyst. Nature Communications 12, 2612 (2021).
7. Kim, Y., Creel, E. B., Corson, E. R., McCloskey, B. D., Urban, J. J. & Kostecki, R., Surface-Plasmon-Assisted Photoelectrochemical Reduction of CO2 and NO3- on Nanostructured Silver Electrodes. Advanced Energy Materials 8, 1800363 (2018).
8. Wan, L., Zhou, Q., Wang, X., Wood, T. E., Wang, L., Duchesne, P. N., Guo, J., Yan, X., Xia, M., Li, Y. F. & others, Cu2O nanocubes with mixed oxidation-state facets for (photo) catalytic hydrogenation of carbon dioxide. Nature Catalysis 2, 889–898 (2019).
9. Iyengar, P., Kolb, M. J., Pankhurst, J. R., Calle-Vallejo, F. & Buonsanti, R., Elucidating the facet-dependent selectivity for CO2 electroreduction to ethanol of Cu–Ag tandem catalysts. Acs Catalysis 11, 4456–4463 (2021).
10. Chen, C., Li, Y., Yu, S., Louisia, S., Jin, J., Li, M., Ross, M. B. & Yang, P., Cu-Ag tandem catalysts for high-rate CO2 electrolysis toward multicarbons. Joule 4, 1688–1699 (2020).
11. Chang, Z., Huo, S., Zhang, W., Fang, J. & Wang, H., The tunable and highly selective reduction products on Ag@ Cu bimetallic catalysts toward CO2 electrochemical reduction reaction. The Journal of Physical Chemistry C 121, 11368–11379 (2017).
12. Landaeta, E., Kadosh, N. I. & Schultz, Z. D., Mechanistic study of plasmon-assisted in situ photoelectrochemical CO2 reduction to acetate with a Ag/Cu2O nanodendrite electrode. ACS Catalysis 13, 1638–1648 (2023).
13. Zhang, T., Liu, Y., Yang, C., Tian, L., Yan, Y. & Wang, G., Monotonically increasing relationship between conversion selectivity from CO2 to CO and the interface area of Cu-Ag biphasic electrochemical catalyst. Journal of Alloys and Compounds 947, 169638 (2023).
14. Choukroun, D., Pacquets, L., Li, C., Hoekx, S., Arnouts, S., Baert, K., Hauffman, T., Bals, S. & Breugelmans, T., Mapping composition–selectivity relationships of supported sub-10 nm Cu–Ag nanocrystals for high-rate CO2 electroreduction. ACS nano 15, 14858–14872 (2021).
15. Clark, E. L., Hahn, C., Jaramillo, T. F. & Bell, A. T., Electrochemical CO2 reduction over compressively strained CuAg surface alloys with enhanced multi-carbon oxygenate selectivity. Journal of the American Chemical Society 139, 15848–15857 (2017).

16. Wang, H., Zhou, X., Yu, T., Lu, X., Qian, L., Liu, P. & Lei, P., Surface restructuring in AgCu single-atom alloy catalyst and self-enhanced selectivity toward CO2 reduction. Electrochimica Acta 426, 140774 (2022).
17. Xu, Y., Li, C., Xiao, Y., Wu, C., Li, Y., Li, Y., Han, J., Liu, Q. & He, J., Tuning the selectivity of liquid products of CO2RR by Cu–Ag alloying. ACS Applied Materials & Interfaces 14, 11567–11574 (2022).
18. Lv, X., Shang, L., Zhou, S., Li, S., Wang, Y., Wang, Z., Sham, T.-K., Peng, C. & Zheng, G., Electron-deficient Cu sites on Cu3Ag1 catalyst promoting CO2 electroreduction to alcohols. Advanced Energy Materials 10, 2001987 (2020).
19. Du, C., Mills, J. P., Yohannes, A. G., Wei, W., Wang, L., Lu, S., Lian, J.-X., Wang, M., Guo, T., Wang, X. & others, Cascade electrocatalysis via AgCu single-atom alloy and Ag nanoparticles in CO2 electroreduction toward multicarbon products. Nature Communications 14, 6142 (2023).
20. Hoang, T. T., Verma, S., Ma, S., Fister, T. T., Timoshenko, J., Frenkel, A. I., Kenis, P. J. A., Gewirth, A. A. Nanoporous copper–silver alloys by additive-controlled electrodeposition for the selective electroreduction of CO2 to ethylene and ethanol. Journal of the American Chemical Society, 140(17), 5791-5797, (2018).
21. Lee, C., Shin, K., Lee, Y. J., Jung, C. & Lee, H. M., Effects of shell thickness on Ag-Cu2O core-shell nanoparticles with bumpy structures for enhancing photocatalytic activity and stability. Catalysis Today 303, 313–319 (2018).
22. Chen, P.-C., Chen, C., Yang, Y., Maulana, A. L., Jin, J., Feijoo, J. & Yang, P., Chemical and structural evolution of AgCu catalysts in electrochemical CO2 reduction. Journal of the American Chemical Society 145, 10116–10125 (2023).
23. Kamat, G. A., Yan, C., Osowiecki, W. T., Moreno-Hernandez, I. A., Ledendecker, M. & Alivisatos, A. P., Self-limiting shell formation in Cu@ Ag Core–shell nanocrystals during galvanic replacement. The Journal of Physical Chemistry Letters 11, 5318–5323 (2020).
24. Muzikansky, A., Nanikashvili, P., Grinblat, J. & Zitoun, D., Ag dewetting in Cu@ Ag monodisperse core–shell nanoparticles. The Journal of Physical Chemistry C 117, 3093–3100 (2013).
25. Osowiecki, W. T., Ye, X., Satish, P., Bustillo, K. C., Clark, E. L. & Alivisatos, A. P., Tailoring morphology of Cu–Ag nanocrescents and core–shell nanocrystals guided by a thermodynamic model. Journal of the American Chemical Society 140, 8569–8577 (2018).
26. Piccinin, S., Stampfl, C. & Scheffler, M., First-principles investigation of Ag-Cu alloy surfaces in an oxidizing environment. Physical Review B—Condensed Matter and Materials Physics 77, 075426 (2008).
27. Schweinar, K., Beeg, S., Hartwig, C., Rajamathi, C. R., Kasian, O., Piccinin, S., Prieto, M. J., Tanase, L. C., Gottlob, D. M., Schmidt, T. & others, Formation of a 2D Meta-stable Oxide by Differential Oxidation of AgCu Alloys. ACS Applied Materials & Interfaces 12, 23595–23605 (2020).
28. Liu, G., Zheng, F., Li, J., Zeng, G., Ye, Y., Larson, D. M., Yano, J., Crumlin, E. J., Ager, J. W., Wang, L.-w. & others, Investigation and mitigation of degradation mechanisms in Cu2O photoelectrodes for CO2 reduction to ethylene. Nature Energy 6, 1124–1132 (2021).
29. Dal Forno, S., Ranno, L. & Lischner, J., Material, size, and environment dependence of plasmon-induced hot carriers in metallic nanoparticles. The Journal of Physical Chemistry C 122, 8517–8527 (2018).
30. Seemala, B., Therrien, A. J., Lou, M., Li, K., Finzel, J. P., Qi, J., Nordlander, P. & Christopher, P., Plasmon-mediated catalytic O2 dissociation on Ag nanostructures: hot electrons or near fields? ACS Energy Letters 4, 1803–1809 (2019).
31. Kumari, G., Zhang, X., Devasia, D., Heo, J. & Jain, P. K., Watching visible light-driven CO2 reduction on a plasmonic nanoparticle catalyst. ACS nano 12, 8330–8340 (2018).
32. Khalakhan, I., Vorokhta, M., Xie, X., Piliai, L. & Matolínová, I., On the interpretation of X-ray photoelectron spectra of Pt-Cu bimetallic alloys. Journal of Electron Spectroscopy and Related Phenomena 246, 147027 (2021).
33. Vasquez, R. P., CuO by XPS. Surface Science Spectra 5, 262–266 (1998).
34. Vasquez, R. P., Cu (OH) 2 by XPS. Surface Science Spectra 5, 267–272 (1998).
35. Hoflund, G. B., Hazos, Z. F. & Salaita, G. N., Surface characterization study of Ag, AgO, and Ag 2 O using x-ray photoelectron spectroscopy and electron energy-loss spectroscopy. Physical Review B 62, 11126 (2000).
36. Deng, X., Verdaguer, A., Herranz, T., Weis, C., Bluhm, H. & Salmeron, M., Surface chemistry of Cu in the presence of CO2 and H2O. Langmuir 24, 9474–9478 (2008).
37. Espinós, J. P., Morales, J., Barranco, A., Caballero, A., Holgado, J. P. & González-Elipe, A. R., Interface effects for Cu, CuO, and Cu2O deposited on SiO2 and ZrO2. XPS determination of the valence state of copper in Cu/SiO2 and Cu/ZrO2 catalysts. The Journal of Physical Chemistry B 106, 6921–6929 (2002).
38. Smekal, W., Werner, W. S., & Powell, C. J. Simulation of electron spectra for surface analysis (SESSA): a novel software tool for quantitative Auger-electron spectroscopy and X-ray photoelectron spectroscopy. Surface and Interface Analysis: An International Journal devoted to the development and application of techniques for the analysis of surfaces, interfaces and thin films, 37(11), 1059-1067 (2005).
39. Ye, Y., Qian, J., Yang, H., Su, H., Lee, K.-J., Etxebarria, A., Cheng, T., Xiao, H., Yano, J., Goddard III, W. A. & others, Synergy between a silver–copper surface alloy composition and carbon dioxide adsorption and activation. ACS applied materials & interfaces 12, 25374–25382 (2020).
40. Gurevich, A. B., Bent, B. E., Teplyakov, A. V. & Chen, J. G., A NEXAFS investigation of the formation and decomposition of CuO and Cu2O thin films on Cu (100). Surface science 442, L971–L976 (1999).
41. Velasco-Vélez, J.-J., Jones, T., Gao, D., Carbonio, E., Arrigo, R., Hsu, C.-J., Huang, Y.-C., Dong, C.-L., Chen, J.-M., Lee, J.-F. & others, The role of the copper oxidation state in the electrocatalytic reduction of CO2 into valuable hydrocarbons. ACS Sustainable Chemistry & Engineering 7, 1485–1492 (2018).
42. Eom, N., Messing, M. E., Johansson, J. & Deppert, K., General trends in core–shell preferences for bimetallic nanoparticles. ACS nano 15, 8883–8895 (2021).

43. Matte, L. P., Thill, A. S., Lobato, F. O., Novôa, M. T., Muniz, A. R., Poletto, F. & Bernardi, F., Reduction-Driven 3D to 2D Transformation of Cu Nanoparticles. Small 18, 2106583 (2022).
44. Breßler, I., Kohlbrecher, J. & Thünemann, A. F., SASfit: a tool for small-angle scattering data analysis using a library of analytical expressions. Applied Crystallography 48, 1587–1598 (2015).
45. Boucher, B., Chieux, P., Convert, P. & Tournarie, M., Small-angle neutron scattering determination of medium and long range order in the amorphous metallic alloy TbCu3. 54. Journal of Physics F: Metal Physics 13, 1339 (1983).
46. Pospelov, G., Van Herck, W., Burle, J., Carmona Loaiza, J. M., Durniak, C., Fisher, J. M., Ganeva, M., Yurov, D. & Wuttke, J., BornAgain: software for simulating and fitting grazing-incidence small-angle scattering. Applied Crystallography 53, 262–276 (2020).
47. Henke, B. L., Gullikson, E. M. & Davis, J. C., X-ray interactions: photoabsorption, scattering, transmission, and reflection at E= 50-30,000 eV, Z= 1-92. Atomic data and nuclear data tables 54, 181–342 (1993).
48. Jaugstetter, M., Qi, X., Chan, E. M., Salmeron, M., Wilson, K. R., Nemšák, S. & Bluhm, H., Direct Observation of Morphological and Chemical Changes during the Oxidation of Model Inorganic Ligand-Capped Particles. ACS nano 19, 418–426 (2024).
49. Krycka, K. L., Borchers, J. A., Salazar-Alvarez, G., López-Ortega, A., Estrader, M., Estradé, S., Winkler, E., Zysler, R. D., Sort, J., Peiró, F. & others, Resolving material-specific structures within Fe3O4 γ-Mn2O3 core shell nanoparticles using anomalous small-angle X-ray scattering. ACS nano 7, 921–931 (2013).
50. Matte, L. P., Jaugstetter, M., Thill, A. S., Mishra, T. P., Escudero, C., Conti, G., Poletto, F., Nemsak, S. & Bernardi, F., Direct observation of phase change accommodating hydrogen uptake in bimetallic nanoparticles. ACS nano (2025).
51. Wu, C. H., Liu, C., Su, D., Xin, H. L., Fang, H. T., Eren, B., Zhang, S., Murray, C. B., Salmeron, M. B. Bimetallic synergy in cobalt–palladium nanocatalysts for CO oxidation. Nature Catalysis, 2(1), 78-85 (2019).
52. Chang, C.-J., Lin, S.-C., Chen, H.-C., Wang, J., Zheng, K. J., Zhu, Y. & Chen, H. M., Dynamic reoxidation/reduction-driven atomic interdiffusion for highly selective CO2 reduction toward methane. Journal of the American Chemical Society 142, 12119–12132 (2020).
53. Liu, G., Lee, M., Kwon, S., Zeng, G., Eichhorn, J., Buckley, A. K., Toste, F. D., Goddard III, W. A., Toma, F. M. CO2 reduction on pure Cu produces only H2 after subsurface O is depleted: Theory and experiment. Proceedings of the National Academy of Sciences, 118(23), e2012649118 (2021).
54. Grass, M. E., Karlsson, P. G., Aksoy, F., Lundqvist, M., Wannberg, B., Mun, B. S., Hussain, Z. & Liu, Z., New ambient pressure photoemission endstation at Advanced Light Source beamline 9.3. 2. Review of Scientific Instruments 81 (2010).
55. Kersell, H., Chen, P., Martins, H., Lu, Q., Brausse, F., Liu, B.-H., Blum, M., Roy, S., Rude, B., Kilcoyne, A. & others, Simultaneous ambient pressure X-ray photoelectron spectroscopy and grazing incidence X-ray scattering in gas environments. Review of scientific instruments 92 (2021).
56. Kersell, H., Weber, M. L., Falling, L., Lu, Q., Baeumer, C., Shirato, N., Rose, V., Lenser, C., Gunkel, F. & Nemšák, S., Evolution of surface and sub-surface morphology and chemical state of exsolved Ni nanoparticles. Faraday discussions 236, 141–156 (2022).
57. Schneider, C. A., Rasband, W. S. & Eliceiri, K. W., NIH Image to ImageJ: 25 years of image analysis. Nature methods 9, 671–675 (2012).
58. Thompson, A. P., Aktulga, H. M., Berger, R., Bolintineanu, D. S., Brown, W. M., Crozier, P. S., In't Veld, P. J., Kohlmeyer, A., Moore, S. G., Nguyen, T. D. & others, LAMMPS-a flexible simulation tool for particle-based materials modeling at the atomic, meso, and continuum scales. Computer physics communications 271, 108171 (2022).
59. Daw, M. S., Foiles, S. M. & Baskes, M. I., The embedded-atom method: a review of theory and applications. Materials Science Reports 9, 251–310 (1993).
60. Williams, P. L., Mishin, Y. & Hamilton, J. C., An embedded-atom potential for the Cu–Ag system. Modelling and Simulation in Materials Science and Engineering 14, 817 (2006).

**Acknowledgements**

This study was funded by FAPERGS (24/2551-0002160-4, and 24/2551-0001952-9), CNPq (310142/2021-0), and CAPES (Finance Code 001). This research used APPEXS endstation at beamlines 9.3.2 and 11.0.2.1 of the Advanced Light Source, which is a DOE Office of Science User Facility under contract no. DE-AC02-05CH11231. Work at the Molecular Foundry was supported by the Office of Science, Office of Basic Energy Sciences, of the U.S. Department of Energy under Contract no. DE-AC02-05CH11231. G. Z. G., and F. B. thank the CNPq for the research grant. G. Z. G. thanks CAPES for the research grant. M. J. and M. S. were supported by the Catalysis program FWP CH030201. The authors also thank CEOMAT-UFRGS, LAPOL-UFRGS, and CM-UFMG staff for their assistance. D. K. was supported in part by an ALS Collaborative Postdoctoral Fellowship.